\documentclass{sigchi-ext}
\usepackage[T1]{fontenc}
\usepackage{textcomp}
\usepackage[scaled=.92]{helvet} % for proper fonts
\usepackage{graphicx} % for EPS use the graphics package instead
\usepackage{balance}  % for useful for balancing the last columns
\usepackage{booktabs} % for pretty table rules
\usepackage{ccicons}  % for Creative Commons citation icons
\usepackage{ragged2e} % for tighter hyphenation

\def\plaintitle{Creative Use of XAI In Socio-Technical Systems: A Case Study} 
\def\emptyauthor{}
\def\plainkeywords{Artificial Intelligence, Explainability, Human-Centered Approaches, Trust, Insurance}

\title{Creative Uses of AI Systems and their Explanations: A Case Study from Insurance}
\numberofauthors{6}
\author{%
  \alignauthor{%
    \textbf{Michaela Benk} \\
    \affaddr{ETH Zurich} \\
    \affaddr{Chair of Technology Marketing} \\
    \affaddr{Mobiliar Lab for Analytics} \\
    \affaddr{Zurich, Switzerland}\\
    \email{mbenk@ethz.ch} } \vfil \alignauthor{%
    \textbf{Raphael P. Weibel}\\
    \affaddr{ETH Zurich}\\
    \affaddr{Chair of Technology Marketing} \\
    \affaddr{Mobiliar Lab for Analytics}\\
    \affaddr{Zurich, Switzerland}\\
    \email{raweibel@ethz.ch} } \vfil \alignauthor{%
    \textbf{Andrea Ferrario}\\
    \affaddr{ETH Zurich} \\
    \affaddr{Chair of Technology Marketing} \\
    \affaddr{Mobiliar Lab for Analytics}\\
    \affaddr{Zurich, Switzerland} \\
    \email{aferrario@ethz.ch} }}

\definecolor{linkColor}{RGB}{6,125,233}
\hypersetup{%
  pdftitle={\plaintitle},
  pdfauthor={\emptyauthor},
  pdfkeywords={\plainkeywords},
  bookmarksnumbered,
  pdfstartview={FitH},
  colorlinks,
  citecolor=black,
  filecolor=black,
  linkcolor=black,
  urlcolor=linkColor,
  breaklinks=true,
}

\begin{document}

%% For the camera ready, use the commands provided by the ACM in the Permission Release Form.
\CopyrightYear{2022}
\setcopyright{rightsretained}
\conferenceinfo{ACM CHI 2022 Workshop on Human-Centered Explainable AI (HCXAI)}{}
\isbn{978-1-4503-6819-3/20/04}
\doi{https://doi.org/10.1145/3334480.XXXXXXX}
%% Then override the default copyright message with the \acmcopyright command.
\copyrightinfo{\acmcopyright}

\maketitle

% Uncomment to disable hyphenation (not recommended)
% https://twitter.com/anjirokhan/status/546046683331973120
\RaggedRight{} 

% Do not change the page size or page settings.
\begin{abstract}
Recent works have recognized the need for human-centered perspectives when designing and evaluating human-AI interactions and explainable AI methods. Yet, current approaches fall short at intercepting and managing unexpected user behavior resulting from the interaction with AI systems and explainability methods of different stakeholder groups. In this work, we explore the use of AI and explainability methods in the insurance domain. In an qualitative case study with participants with different roles and professional backgrounds, we show that AI and explainability methods are used in creative ways in daily workflows, resulting in a divergence between their intended and actual use. Finally, we discuss some recommendations for the design of human-AI interactions and explainable AI methods to manage the risks and harness the potential of unexpected user behavior.
%Recent works have recognized the need for human-centered perspectives when evaluating XAI, yet current approaches presume specific and static processes and roles for users and AIs. In this work, we explore the dynamic of AI and explainability in the insurance domain. By interviewing members of multiple teams and two different stakeholder groups, we show that when AIs are embedded in socio-technical systems, it is not always clear how different users will interact with and use XAI. Our findings suggest that AI tools and explainability methods are used in creative ways, thus creating divergence between intended and actual use. We discuss takeaways and potential pitfalls, as well as offer recommendations for future work.
\end{abstract}


% ACM Classfication



% Print the classficiation codes
%Please use the 2012 Classifiers and see this link to embed them in the text: \url{https://dl.acm.org/ccs/ccs_flat.cfm}

%%%%%%%%%%%%%%%%%%%%%%%%%%%%%%%%%%%%%%%%%%%%%%%%%%%%%%%%%%%%
%%%%%%%%%%%%%%%%%%%%%%%%%%%%%%%%%%%%%%%%%%%%%%%%%%%%%%%%%%%%
\section{Introduction}
The field of explainable AI (XAI) has produced numerous methods to provide insights into ML-based systems and to promote transparency, understanding, and trust \cite{Ribeiro-et-al_2016, miller_explanation_2018, Langer2021WhatDW}. As ``one explanation does not fit all'' \cite{Arya2019OneED}, prior work has considered different use cases and stakeholder groups of XAI \cite{Arya2019OneED, sun2022investigating, Liao2020QuestioningTA, Bhatt2020ExplainableML}. Findings suggest that XAI methods are often used as ``sanity checks'' for users in specific roles, such as data scientists \cite{Bhatt2020ExplainableML}. However, their applicability and appropriate use for other stakeholder groups remains an open research challenge \cite{Knowles2021TheSO}.
%Moreover, only few works have investigated their use for other stakeholder groups, such as end-users \cite{sun2022investigating, Liao2020QuestioningTA}.
% We argue that current XAI methods and their evaluation approaches often limit their scope to users in specific roles, such as a single decision maker probing the AI with questions. Such roles often resemble a ``supervisory control'' setting \cite{Chiou2021TrustingAD}, where one user takes decisions and one AI assists. 

As a result, the current design of human-AI interactions and XAI methods may fall short at intercepting the socio-technical complexity of systems where humans interact with AIs depending on their roles, skills, and expectations. 

\begin{marginfigure}[1pc]
\begin{minipage}{\marginparwidth}
\includegraphics[width=1.0\marginparwidth]{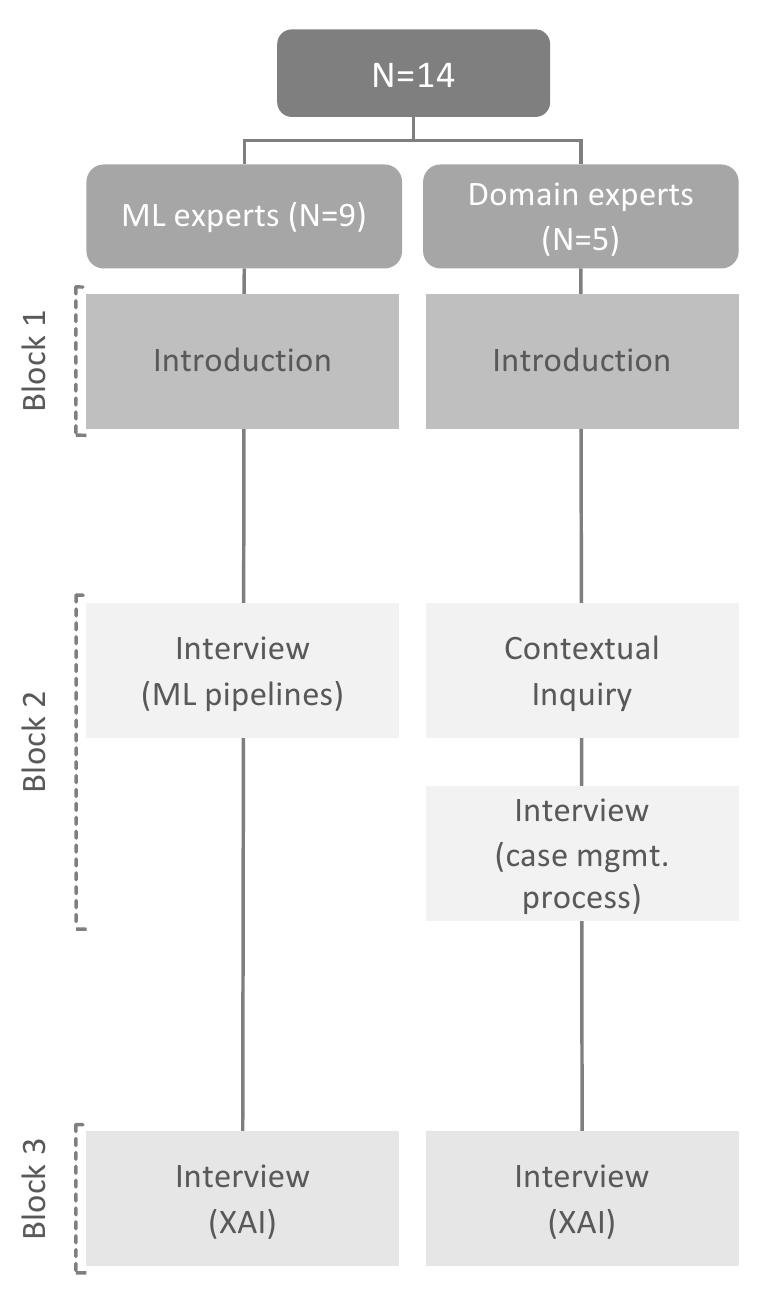}
\caption{Flowchart of the performed assessments.}
\label{fig:flow}
\end{minipage}
\end{marginfigure}
%in different roles of AIs and humans in joint decision making processes, e.g., involving engineers, business owners, and customers \cite{van2020embedding,kroes2006treating}, may require different evaluation approaches. 
In particular, this limitation may result in the emergence of unexpected or ``creative'' uses of AI and XAI methods that deviate from those encoded in their original design. These unexpected uses may be driven by users' specific mental models on the AI and the explanations of its predictions, or by pragmatic reasons, such as the necessity to finalize work tasks on time. In this work, we present the results of a qualitative case study and show that employees, namely machine learning (ML) and domain experts, of an insurance company use AI and XAI methods in creative ways in their daily workflow. These creative uses are driven by different motivations, leading to different trusting relations to manage the challenges and risks arising from human-AI interactions in a professional setting. 
%By focusing on the data intensive and highly process-driven insurance domain, we could gain insights into business processes and explainability needs at different stages of the decision process and from multiple perspectives, including ML experts and lay users. 
%, we synthesize the results into main takeaways, thereby answering the following research question: What user goals should XAI aim to support in the insurance domain, for whom, and why? % Our contribution is twofold. First, we present the results of interviews of two main stakeholder groups and several teams, conducted at a large Swiss insurance company. 
% the observed use of XAI methods and the interactions with AI may diverge from their intended use and explores novel ways. \textcolor{red}{Recommendation lines}
%We thereby hope to inform future work aimed at designing and investigating XAI in socio-technical systems.%supporting such socio-technical settings.
%We thereby hope to inform future work, by advocating for a more holistic and context-specific approach to the development and evaluation of XAI ...

%%%%%%%%%%%%%%%%%%%%%%%%%%%%%%%%%%%%%%%%%%%%%%%%%%%%%%%%%%%%%%%%%%%%
%%%%%%%%%%%%%%%%%%%%%%%%%%%%%%%%%%%%%%%%%%%%%%%%%%%%%%%%%%%%%%%%%%%%
\section{Methodology}

%\subsection{\textbf{Interviews}}
We conducted two separate sets of assessments comprising semi-structured interviews and a contextual inquiry with N=14 employees of a Swiss insurance company. In the first set, we interviewed nine employees whose tasks involve ML modeling (henceforth referred to as ``ML experts''). Specifically, they included two pricing actuaries\footnote{In Switzerland, pricing actuaries are data scientists or statisticians who specialize in either the financial or insurance industries with a focus on the quantitative pricing of products and services.}, one data engineer, and six data scientists.
%of which two were tasked with pricing, two with customer analytics, one with fraud detection and one with reporting. 
%Their time of employment with the insurance company ranged from one month to five years. 

In the second set, we interviewed five employees using an AI that classifies insurance claims (e.g., household or car damage) by means of ML and natural language processing (NLP) on textual user inputs. They included four insurance case handlers and one case manager (henceforth referred to as ``domain experts''). 
%Similar to the first group their time of employment ranged from 1.5 years to 9 years.
%The structure of our protocol was inspired by the work of Kaur et al. \cite{kaur-et-al_2020}. 
Both sets of assessment were structured in three blocks, as seen in Figure \ref{fig:flow}. The first block (ca. 10 minutes) consisted of questions to record the participants' demographics and background (e.g., their current role and professional tasks). The second block (ca. 20 minutes) focused on their use of AI systems. For the first set (ML experts), we asked semi-structured interview questions focusing on the ML pipelines they use in their workflow and their challenges. For the second set (domain experts), we observed how participants used the AI with a contextual inquiry (ca. 20 minutes), before continuing with semi-structured interviews (ca. 20 minutes) to understand the case management process in more detail. In the third and last block, we continued with semi-structured interview questions, focusing on the use of XAI methods, such as explanations, for both sets of experts. We adapted the interview protocol used by Kaur et al. \cite{kaur-et-al_2020} to ask questions, such as ``When are you satisfied with the results [of the model]?'' or ``Would you consider it useful to have an explanation of the [AI] tool’s decision or inner working?''\\
Due to COVID-19 measures, all assessments took place remotely, via video-conference. They were video- and audio-recorded, and subsequently transcribed. We coded and analyzed the assessments using the framework method \cite{ritchie2013qualitative}.
%\subsection{\textbf{Analysis}}

%%%%%%%%%%%%%%%%%%%%%%%%%%%%%%%%%%%%%%%%%%%%%%%%%%%%%%%%%%
%%%%%%%%%%%%%%%%%%%%%%%%%%%%%%%%%%%%%%%%%%%%%%%%%%%%%%%%%%
\section{Findings}
% Based on the framework method and the interview structure, we grouped findings into three over-arching themes. 
Based on the framework method, we identified the following themes: 1) use of AI and XAI methods, 2) trusting processes, and 3) challenges and risks in using AI in insurance processes. 
% Below we elaborate on each theme, with respect to the two stakeholder groups.
%challenges and risks with respect to users' workflow

%%%%%%%%%%%%%%%%%%%%%%%%%%%%%%%%%%%%%%%%%%%%%%%%%%%%%%%%%%%%%%
\subsection{\textbf{ML experts}}

\noindent{\textit{(X)AI use.}} All participants considered explainability an overall important aspect of their work. Most of them emphasized that they prefer or are required to use interpretable models. This is driven, we argue, by the modeling requirements that are in place in the actuarial practice. One participant noted: ``I feel more comfortable with [...] the Generalized Linear Model.[...] We can clearly see what is the prediction, what was the observed value for this feature, what is the exposure for each level, and so on.''
Furthermore, explanations are deemed necessary for communication with other stakeholders. For instance, one participant explained: 
%``The SHAP analysis...it's really a visual and very comprehensive tool to help us explain to business how certain features influence other features.'' 
``A black box in industry is somehow difficult. I mean, people need to understand, what is happening [...]. They do not want to know [the inner workings], but instead, 'okay, does this make sense?'''.
Interestingly, explanations are also used in creative ways. In fact, instead of using them to ``classically" validate and debug a given model \cite{watson2021explanation}, some participants employ a two-step approach to XAI. 
% Recognizing the value of high performing models in generating insights into new potential features, 
First, they use SHAP values (SHapley Additive exPlanations) \cite{Lundberg_2017} to explain feature interactions generated by a high performing, yet complex, model (e.g., a gradient boosting machine). Then, they use these insights to augment the \textit{original} interpretable, yet simpler, model. 
% Interpretability important, black boxes are hard to sell to non-experts, but if performance is good, it's easier (depending on use case)
% Performance vs Interpretability trade off: mixed results some prefer one some the oter, some say based on case by case (further evidence for human-centered with context approach by Liao)
% - Interpretability is considered important
% - Use of XAI limited to development. 
% - XAi could be helpful when communicating to "business" stakeholders (decision makers)
% - Users of ML systems would benefit from explanations (by reducing barrier to entry, easier adaption)

\noindent{\textit{Trust.}} ML experts frequently mentioned explainability as being relevant to enable trust. However, they mainly considered trust relevant when communicating with non-experts. As noted by one participant: ``I find it important that the people [non-experts] understand what happens [with the model] and can picture it.'' When evaluating a ML model output, they claimed to use also their ``gut feeling'', which they built through familiarity with and knowledge of ML models and the use of different performance measures.
%However, it was less of an issue for ML experts themselves, and more of an issue with respect to non-experts. As noted by one participant: ``I find it important that the people [non-experts] believe in what happens [with the model] and can picture it.'' 

%For participants ML pipeline, familiarity with and knowledge of the models and 
%performance-related metrics re-assured them and allowed them to build a ``gut feeling'' of the output. 
%When asked about trust in SHAP's output, one participant highlighted: ``Well, for me it is always clear, as I understand the logic behind it and checked the source code.'' %uncertainty reduction

\noindent{\textit{Challenges and Risks.}} The main challenges highlighted by ML experts included adequate data quality and processing, as well as communication with other stakeholders. In fact, for most ML experts, the retrieval of adequate amounts of high quality data was a critical issue for ML modeling. Moreover, they mentioned that internal communication is challenging as it requires translating ML concepts into a language that can be understood by non-experts (a challenge one participant referred to as ``[providing] explanations of the explanations'').

%%%%%%%%%%%%%%%%%%%%%%%%%%%%%%%%%%%%%%%%%%%%%%%%%%%%%%%%%%%%%%%%
%%%%%%%%%%%%%%%%%%%%%%%%%%%%%%%%%%%%%%%%%%%%%%%%%%%%%%%%%%%%%%%%
\subsection{\textbf{Domain experts}}

\noindent{\textit{(X)AI use.}} All but one participant were largely unaware of the use of ML and NLP-based text classification and believed the AI to be based on keyword matching. As a result, some participants \textit{actively} adjusted their writing style in accordance with their mental model of the AI capabilities. For instance, some included specific keywords during the interactions with the AI, as they believed this would have affected the quality of the AI predictions. One employee noted ``It helps the system, if one writes the keyword first and then the description [of the claim],'' and another: ``The system is at times confused, when one writes too much.'' 
Others were less ``creative'' and opted for a more process-oriented approach in their interaction with the AI. To do so, they included as much textual information on the claim as possible in the user interface to facilitate the subsequent management of the case by claim experts.

% Participants were asked whether it would help their work if the algorithm provided an explanation as to why a certain claim class was proposed. 
During the interviews, we showed two pictures of possible XAI explanations (feature highlighting or counterfactual explanation) to participants, to support their understanding of what an explanation might look like. All participants indicated that they would appreciate the provision of explanations of the AI predictions. However, some specified that, due to time constraints, their use would not be feasible for all cases at hand. In particular, one participant specified that an explanation would mostly be interesting out of curiosity. Another noted that it would particularly benefit new users who lack familiarity with the AI.    
%Another participant was highly interested in such an addition to the tool and mentioned that it would be of particular help to see an illustration (i.e. a decision tree) of how the algorithm reached its decision. 
%Some participants were curious about the system’s logic and paid attention to changes to input, which affect the output. One participant ``played around'' with the text input in order to better understand the system’s logic and stated that it was mostly due to curiosity. Furthermore, one participant appreciated the ability to ``learn from the system.''

\noindent{\textit{Trust.}} Overall, participants mentioned that they have trust in the \textit{process}, rather than in the AI. To this end, one participant stated that they do not trust the AI at all, but they trust their colleagues to wrap-up the case correctly. %Another mentioned: ``It is no problem if the [AI] output is wrong, as my colleagues will adjust it.''
%One participant trusted the tool and attributed mistakes to a lack of knowledge of how to use it; particularly the confirmation questions posed by the tool ``It appears the case you described is case X -- is this correct?'' contributed to added confidence in the classification. 

\noindent{\textit{Challenges and risks.}} Participants mentioned that the use of the AI could potentially affect the quality of the claims process (e.g., its timely completion and the payouts), and the communications with customers. On the one hand, the process in which the AI is embedded offers a degree of protection against the AI errors: ``It is no problem if the output is wrong, as the claim managers will adjust it.'' On the other hand, two participants emphasized that any incorrect AI outputs should not be shared with customers, as discrepancies between the initial information provided over the phone and the continued process would be highly undesirable. 
%
%Two case handlers did not perceive the correctness of the claim classification as overly important, as the claims team would be able to make necessary changes to any mistakes in a registered claim. One person mentioned: ``It is no problem if the output is wrong, as the claim managers will adjust it.''
%In contrast, a claim manager mentioned that a wrongly classified claim requires deletion and re-submission, necessitating additional time and effort for managers. 
% Two participants emphasized the importance of correctness of any system output potentially shared with the client, as discrepancies between the initial information provided over the phone and the continued process would be highly undesirable. 
%
%  Central to the supervisor-control approach is the focus on trust calibration, which assumes a human requires information on the AIs inner workings or outcome generation, in order to assess its trustworthiness. However, we have found that...

%%%%%%%%%%%%%%%%%%%%%%%%%%%%%%%%%%%%%%%%%%%%%%%%%%%%%%%%%%%%%
%%%%%%%%%%%%%%%%%%%%%%%%%%%%%%%%%%%%%%%%%%%%%%%%%%%%%%%%%%%%%
\section{Recommendations for Designers}

The creative use of AI and XAI methods is a double-edged sword in the hand of the designers of human-AI interactions. On the one hand, unexpected uses may reveal unforeseen design possibilities that can be implemented at a later stage. This may lead to an improvement of the interactions with respect to their original design. On the other hand, creative uses may represent shortcuts that users take to address specific limitations in the human-AI interaction, leading to the misuse of the systems and the explanations of their predictions. This, in turn, may eventually affect the willingness to use and trust in the AI. 

In summary, the emergence of creative uses of AI and XAI methods is a possibility that designers may need to consider and harness to improve human-AI interactions in real-world applications.\\ 

Our study shows that ML experts may use XAI methods (e.g., SHAP values) on high-performing and complex ML models to improve the performance on an interpretable and simpler model.
These methods are seen as a support to trust in the AI. Therefore, designers of human-AI interactions for ML experts may promote a two-step approach to XAI in those domains where the use of these methods is emerging and the implementation of interpretable models is seen as a necessity, due to, for example, specific business values or requirements (e.g., regulatory constraints).

\noindent{\textbf{Takeaway:}} Users may interact with AI systems and their explanations in ways that are not anticipated by designers as a result of business values and specific requirements, such as promoting interpretable and simple models.\\ 
\noindent{\textbf{Research Opportunity:}} IInvestigate the trade-off between the AI system flexibility to support its creative use and its complexity that may lead to its unintended use.

In addition, we have shown that domain experts may actively try to affect the qualities of the AI predictions by adjusting their writing style while interacting with the system to classify an open case of insurance claim. This creative interaction with the AI is, arguably, the result of some users' wrong mental model of the system functionalities (i.e., that the AI predictions are computed based on keywords instead of NLP on free text). This could in turn have an effect on downstream processes, e.g., affecting the quality of data generation for future ML training routines. 
Here, designers could consider the use of methods (e.g., textual cues, conversational agents) to manage the risk of misuse of the system resulting from incorrect users' mental models and support the efficiency of the process in which the human-AI interaction takes place. 

Here, designers could consider the use of methods (e.g., textual cues, conversational agents) to manage the risk of misuse of the system resulting from incorrect users’ mental models and support the efficiency of the process in which the human-AI interaction takes place. Moreover, it may be crucial to adapt a process-oriented approach to explainability, as proposed by Ehsan et al., to avoid perpetuating users’ wrong mental models and their updates with explanations of internal mechanisms of the model \cite{Ehsan2021ExpandingET}. For example, in the case of the domain experts of our study, feature highlighting methods may further fuel the belief that the AI predictions are keyword-based. Instead, explanations could be used to make sense of situations, correcting or anticipating user’s knowledge gaps and relating their activities to and building awareness of other users’ activities and their changes over time \cite{Ehsan2021ExpandingET}.

\noindent{\textbf{Takeaway:}} Creative use of AI systems and their explanations may have detrimental consequences on downstream real-world processes. \\
\noindent{\textbf{Research Opportunity:}} Investigate necessary procedural meta-information that explanations should transmit to manage the risks of creative interactions over time.
 
% For ML users, our main finding suggests that users trusted the process that the AI was embedded in, while being largely unaware of the ``AI'' in their tool. 
% This may confirm the premise by \cite{Knowles2021TheSO} that end-users are not actually \textit{users of} AI, but rather \textit{subjects to} AI \cite{Knowles2021TheSO}. 
% Users furthermore adjusted their writing style due to wrong mental models of the system. In this case, explanations, such as feature highlighting, may be misinterpreted and further perpetuate the wrong assumption that the system is based on keyword recognition (while indeed, as conveyed by the engineers, more textual input is needed for correct claim classification).


\end{document}